\documentstyle[prl,multicol,aps,epsfig]{revtex}
\begin{document}
\renewcommand{\textfraction}{0.10} \renewcommand{\topfraction}{1.0}
\renewcommand{\bottomfraction}{1.0} \flushbottom
\title{Ground State Quantum Computation}
\author{Ari Mizel, M. W. Mitchell, and Marvin L. Cohen}
\address{Department of Physics, University of California at Berkeley,
Berkeley, CA 94720, USA, \\ and Materials Sciences Division, Lawrence
Berkeley National Laboratory, Berkeley, CA 94720, USA.  }

\date{\today} 
\maketitle

\begin{abstract}
We formulate a novel ground state quantum computation approach that
requires no unitary evolution of qubits in time: the qubits are fixed
in stationary states of the Hamiltonian.  This formulation supplies a
completely time-independent approach to realizing quantum computers.
We give a concrete suggestion for a ground state quantum computer
involving linked quantum dots.
\end{abstract}

\begin{multicols}{2}
The discovery of efficient quantum mechanical factoring and database
searching algorithms has fueled tremendous research interest in the
field of quantum computing\cite{Shor,Grover1,Grover2}.  Much recent
effort has been dedicated to the problem of realizing a quantum
computer in the
laboratory\cite{CiracZoller,Monroe,Gershenfeld,Chuang,Shnirman} .
This problem is technically challenging, in part because it requires
overcoming the decoherence of quantum mechanical variables as a result
of interactions with the environment\cite{DiVincenzo}.  Several
experimental systems have been proposed as candidates for quantum
computers, and progress has been exciting, but fundamental obstacles
still stand in the way of creating viable computers.  In this letter,
we formulate a novel approach to quantum computing that circumvents
the problem of decoherence, which motivates new directions for quantum
computer design.  Our scheme works exclusively with quantum mechanical
ground states, completely obviating the need for time-dependent
control of a system.  While researchers have considered using quantum
mechanical ground states to perform classical computations
\cite{Castagnoli,Tougaw,automata}, the idea of executing quantum algorithms
using a ground state computer is an exciting unexplored possibility.

In traditional quantum computation, one examines the development in
time of a collection of quantum mechanical ``qubits'' under controlled
unitary evolutions\cite{Lloyd}.  Each qubit is a
two-state system, described by inner products of the
quantum mechanical state $\left|\psi(t)\right>$ with basis states
$\left|0\right>$ and $\left|1\right>$ associated with the 0
and 1 bit values.  As an $N$-step quantum computation proceeds, the 0
and 1 states remain fixed.  The state of the system progresses as
$\left|\psi(t_{i})\right> = U_{i} \left|\psi(t_{i-1})\right>$,
$i=1$ to $N$, where $U_{i}$ is a unitary operator.  The progress
of the computation is described by the $2(N+1)$ inner products
$\left<0|\psi(t_i)\right>$ and $\left<1|\psi(t_i)\right>$, which
evolve according to the matrix equation
\begin{equation}
\label{matrixeqn}       
 \left[\begin{array}{c}
                \left<0|\psi(t_i)\right> \\ \left<1|\psi(t_i)\right>
        \end{array}  
        \right] =  U_{i} 
        \left[\begin{array}{c}
                \left<0|\psi(t_{i-1})\right> \\
\left<1|\psi(t_{i-1})\right>
        \end{array}
        \right] .
\end{equation}

In our method, the qubits do not change in time; they are fixed in
their ground states.  The steps in the computation correspond, not to
evolution between time points, but rather to development of the ground
state between connected parts of the Hilbert space.  Here a ``qubit''
is a single system with $2(N+1)$ available states, grouped
into $(N+1)$ two-state subspaces
$\{\left|0_{i}\right>,\left|1_{i}\right>\}$ one for each stage of the
calculation.  The ground state of the qubit is a superposition
containing at least one component of each subspace,
$\left|\psi_{g}\right> = \sum_{i}\left|0_i\right>
\left<0_{i}|\psi_{g}\right> +
\left|1_i\right>\left<1_{i}|\psi_{g}\right>$.  The progress of the
computation is then described in terms of the $2(N+1)$ inner products
$\left<0_{i}|\psi_{g}\right>$ and $\left<1_{i}|\psi_{g}\right>$.  As
demonstrated below, proper choice of the Hamiltonian leads to a
sequence of these inner products according to equation
(\ref{matrixeqn}), in exact analogy to traditional quantum
computation.

Consider the example of a spin 1/2 particle acting as a qubit.
In a traditional quantum computation scheme, the basis states
$\left|0\right>$ and $\left|1\right>$ would correspond to states
$\left|\downarrow\right>$ and $\left|\uparrow\right>$, respectively.
The state of the particle at time $t_{i}$, $\left|\psi(t_{i})\right>$,
would evolve from time step to time step by external manipulation of
the direction of the particle's spin.  Computation steps would leave
the spatial wavefunction of the particle fixed (and irrelevant to the
computation).  In our approach, the spin 1/2 particle might be
contained in the ground state of a box, and this ground state would
evolve only trivially in time.  However, the Hamiltonian controlling
the particles in the box would be designed so that its ground state
$\left|\psi_{g}\right>$ possessed a desired non-trivial relationship
between spin variables and spatial variables.  As a result, the spin
of the ground state would develop in a desired way through real space.
We would set $\left|0_{i} \right> = \left|{\bf x}={\bf x}_i,\downarrow
\right>$ and $\left|1_i\right> = \left|{\bf x}={\bf x}_i,\uparrow
\right>$, where the ${\bf x}_i$ denote successive spatial locations.
The quantum computation would proceed via the development of
$\left<0_i|\psi_{g}\right>$ and $\left<1_i|\psi_{g}\right>$ in space
across the box.

This approach completely eliminates the need for time-dependent 
control of quantum mechanical variables.  Instead, we face the 
different challenges 
involved in designing and realizing a tunable static Hamiltonian.  While
it 
has been shown \cite{Feynman,Peres} that time-dependent quantum 
computation can be performed with a static ``cursor Hamiltonian,'' 
such an approach requires complicated many-particle interactions and 
time-dependent state preparation and measurement.  Our scheme is 
time-independent in both the Hamiltonian and the state of the 
computer, and requires only two-particle interactions.

%
%

\newcommand{\mycaption}[1]{\parbox{3in}{\small #1}}

\begin{figure}[h]
\centerline{\psfig{width=4.25cm,figure=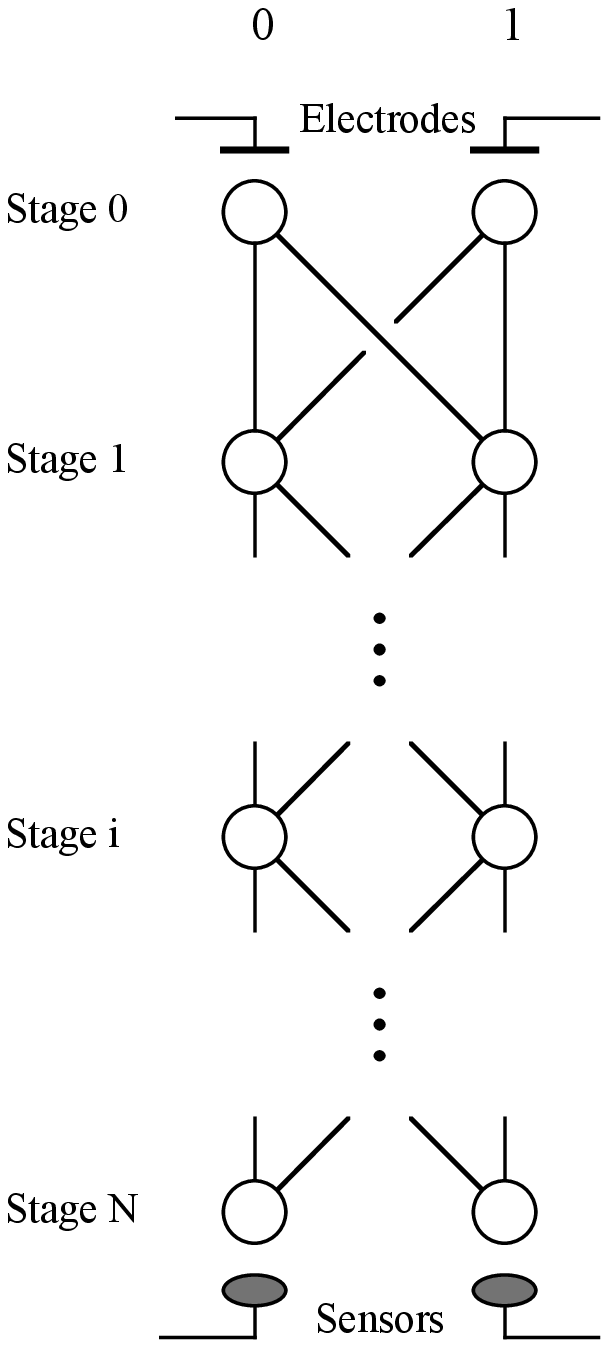}}
\vspace{0.125in}
\mycaption{FIG. 1.  Array of quantum dots that form a single q-bit.  
Lines indicate tunneling paths.}
\label{single}
\end{figure}

We now propose a more detailed implementation of a ground state 
quantum computer in the form of an array of quantum dots.  A single 
qubit is realized by a chain of linked quantum dots, as depicted in 
Fig.  \ref{single}.  At each stage $i$ in the chain, the state 
$\left|0_i\right>$ ($\left|1_i\right>$) describes an electron 
localized on the left (right) dot and represents a 0 (1) value for 
that qubit.  This state can be expressed in terms of a creation 
operator $c^{\dagger}_{i,0}$ ($c^{\dagger}_{i,1}$), where 
$\left|0_i\right> = c^{\dagger}_{i,0}\left| {\rm vac} \right>$ 
($\left|1_i\right> = c^{\dagger}_{i,1}\left| {\rm vac} \right>$).  It 
is convenient to group creation operators together into 
$C^{\dagger}_{i} = \left[ c^{\dagger}_{i,0} , c^{\dagger}_{i,1} 
\right].$

For the quantum computation, we will design the Hamiltonian to have a
twofold degenerate ground state.  In particular, the ground states
$\left|\psi_{g0}\right>$, $\left|\psi_{g1}\right>$ can be established
such that $\left<1_{0}|\psi_{g0}\right> = \left<0_{0}|\psi_{g1}\right>
= 0$, while $\left<0_{0}|\psi_{g0}\right>$ and
$\left<1_{0}|\psi_{g1}\right>$ are nonzero.  These states correspond
to input bit values of 0 and 1, respectively.

To implement a particular computation, $\left|\psi_{g0}\right>$ and
$\left|\psi_{g1}\right>$ can be designed to develop as required by the
algorithm.  The algorithm consists of a sequence of unitary
transformations $U_{i}$, $i=1$ to $N$.  
In the
ground state quantum computer, this transformation requires that the
amplitudes of the components of the ground state $\left|\psi_{g0}\right>$ 
satisfy
\begin{eqnarray} 
\label{transformation}
\left[ \begin{array}{c} \left<0_{i}|\psi_{g0}\right> \\ 
\left<1_{i}|\psi_{g0}\right> \end{array} \right] 
& = & U_{i} \left[ \begin{array}{c} \left<0_{i-1}|\psi_{g0}\right> \\ 
\left<1_{i-1}|\psi_{g0}\right> \end{array} \right] ,
\end{eqnarray}
where $\left<1_{0}|\psi_{g0}\right> = 0$.  For
$\left|\psi_{g1}\right>$ the transformations are analogous but start
with $\left<0_{0}|\psi_{g1}\right> = 0$.  

For what follows, it is convenient to describe how a ground state with 
the desired development can be built up row by row.  Suppose we 
already have a single qubit computer consisting of the $j$ rows $0$ 
to $j-1$ which implements the first $j-1$ unitary operations of an 
algorithm.
Denote the ground
states of this computer by $\left|\psi_{g0}^{j-1} \right>$ and
$\left|\psi_{g1}^{j-1} \right>$.  If we add an additional row $j$ 
(yielding 
$j+1$ total) to the 
computer so that the qubit undergoes an additional unitary operation
$U_{j}$,
then the new ground state $\left|\psi_{g0}^{j} \right>$ is, by 
equation (\ref{transformation})
\begin{equation}
\label{recursion}
\left|\psi_{g0}^{j} \right> =
(1+C^{\dagger}_{j}U_{j}C_{j-1})\left|\psi_{g0}^{j-1} \right>
\end{equation}
and similarly for $\left|\psi_{g1}^{j} \right>$.  Here and 
throughout this letter the uninteresting constant factors which 
maintain normalization of the states have been omitted.

There are many positive semi-definite Hamiltonians that satisfy
$H\left|\psi_{g0}\right> = H\left|\psi_{g1}\right>=0$, thus
possessing the desired degenerate ground states.  For an $N+1$ row
qubit, a particularly convenient Hamiltonian is $H = \sum_{i=1}^{N} 
h^{i}(U_{i})$ where 
\begin{equation}
\label{Hamsingle}
h^{i}(U) \equiv \epsilon
 \left[
 C^{\dagger}_{i-1}C_{i-1} + C^{\dagger}_{i}C_{i} -
 \left(C^{\dagger}_{i} U C_{i-1} + {\rm h.c.}\right)\right]
\end{equation}
and the constant energy $\epsilon$ defines the energy scale of the
Hamiltonian.  This Hamiltonian has an appealing modular character:
the unitary matrix $U_i$ only enters the Hamiltonian through matrix
elements between states on rows $i-1$ and $i$.

The Hamiltonian described thus far has been constructed to have a 
degenerate ground state: one state for each possible input value to 
the algorithm.  The complete Hamiltonian for a calculation with a 
specific input value does not posess this degeneracy.  Rather, the 
on-site energies of the 0th stage quantum dots differ by a 
small amount.  Thus $\left|\psi_{g0}\right>$ or 
$\left|\psi_{g1}\right>$ is the sole ground state, corresponding to an 
input bit-value of 0 or 1, respectively.  Physically, this can be 
accomplished by applying a voltage to one of the electrodes indicated 
in Fig.  1.  With the system in its ground state, sensors 
near the $N$th stage quantum dots measure the results of the 
calculation, i.e., the location of the electron when it is at the 
output stage.

For generality, we have written the elements of the Hamiltonian in 
equation (\ref{Hamsingle}) in terms of arbitrary unitary 
transformations.  For the case of a quantum-dot computer, tunneling 
matrix elements between dots can only be tuned to real values.  (The 
tuning could be accomplished physically by placing intermediate 
quantum dots with controllable voltages in between the primary dots 
shown in Fig. 1.  In this arrangement, the tunneling elements
could 
be changed from run to run in order to implement varied algorithms.)  
Because the tunneling matrix elements will be real, only real unitary, 
i.e., orthogonal, transformations can appear in equation 
(\ref{Hamsingle}).  Other implementions may not have this 
restriction.  As we show below, the restriction does not impede the 
implementation of a nontrivial database search algorithm.  This is 
because the matrix elements are real in important single qubit 
operations like NOT operations
\begin{equation}
\label{NOT}
U = N = \left[\begin{array}{cc} 0 & 1 \\
1 & 0 \end{array} \right]
\end{equation}
and rotations by $\pi/4$ 
\begin{equation}
\label{pioverfour}
U = R(\pi/4) = \left[\begin{array}{cc} \cos \pi/4 & \sin \pi/4 \\
-\sin \pi/4 & \cos \pi/4 \end{array} \right].
\end{equation}
However, for some computations, algorithm
modifications may be necessary.

So far, we have described the behavior of a single qubit quantum 
computer.  A general quantum computer has $M$ qubits each of which 
requires two columns of $(N+1)$ dots for an $N$-step computation.  
Suppose that, at the $j$th stage of the calculation, the algorithm 
specifies the $M$ qubit operation $\cal{U}_{j}$ consisting of 
independent operations for each qubit.  In analogy to 
equation (\ref{recursion}), if $\left| \Psi_{n} ^{j-1}\right>$ is the 
ground state of a $j$ row quantum-dot computer with input $n$ then
\begin{equation}
\label{multinonint}
\left| \Psi_{n}^{j} \right> = \prod _{a=1}^{M} (1 +
C^{\dagger}_{a,j} U_{a,j} C_{a,j-1}) \left| \Psi _{n} ^{j-1}
\right>.
\end{equation}
Here we have introduced an additional index $a$ on each
operator in order to specify to which qubit it
pertains.
This relation has the property that if
$P_{j} \equiv \prod_{a} C^{\dagger}_{a,j}C_{a,j}$ projects
onto the subspace in which all electrons are in the $j$th 
row, then 
\begin{equation}
\label{projectionprop}
P_{j}\left| \Psi _{n} ^{j}\right> = 
{\cal U}_{j}A_{j,j-1}P_{j-1}\left| \Psi_{n}^{j-1}\right>,
\end{equation}
where $A_{j,k}\equiv \prod_{a}C^{\dagger}_{a,j}C_{a,k}$ simply moves 
the electrons from row $k$ to row $j$.
In this way the development of the state reflects the unitary 
evolution of the algorithm.  If the Hamiltonian describing the $j$ row 
computer (which accomplishes $j-1$ operations) is $H^{j-1}$, then the 
Hamiltonian for the $j+1$ row computer is $H^{j} = H^{j-1} + 
\sum_{a=1}^{M} h_{a}^{j}(U_{a,j}).$
    
In addition to single-bit operations, a general quantum computer must 
perform multiple qubit operations such as the controlled-NOT operation 
(CNOT).  Assume the algorithm specifies as the $j$th operation ${\cal 
U}_{j}$ a CNOT of qubit B by qubit A and unitary operations $U_{a,j}$ 
on the other qubits $a\neq A,B$.  Equation (\ref{projectionprop}) 
still holds provided the state of the enlarged array is
\end{multicols}
\begin{equation}
\label{multirecursion}
 \left| \Psi _{n} ^{j}\right> =
\left(1+ c^{\dagger}_{A,j,0}c_{A,j-1,0}(1+C^{\dagger}_{B,j}
C_{B,j-1}) +
c^{\dagger}_{A,j,1}c_{A,j-1,1}(1+C^{\dagger}_{B,j} N C_{B,j-1})\right) 
\prod_{a\ne A,B} 
(1 + C^{\dagger}_{a,j} U_{a,j} C_{a,j-1}) \left|
\Psi _{n} ^{j-1}\right>
\end{equation}
\begin{multicols}{2}
\noindent
where $N$ is the NOT operation matrix (\ref{NOT}).  

This state is the ground state of the enlarged quantum computer if the 
enlarged Hamiltonian is
\begin{equation}
\label{enlargedham}
H^{j} = H^{j-1} + h^{j}_{A,B}({\rm CNOT})
+ \sum_{a=1,a\neq A,B}^{M} h_{a}^{j}(U_{a,j})
\end{equation}
where $h^{j}_{A,B}({\rm CNOT})$ is the addition to the Hamiltonian 
which effects the CNOT operation.
\begin{eqnarray}
\label{reconnect1}
h^{j}_{A,B}({\rm CNOT}) &=& 
\epsilon  C_{A,j-1}^{\dagger}C_{A,j-1}C_{B,j}^{\dagger}C_{B,j} \\
\label{reconnect2}
&&+ h_{A}^{j}(I) C_{B,j-1}^{\dagger}C_{B,j-1}  \\
\label{reconnect3}
&&+ c_{A,j,0}^{\dagger}c_{A,j,0} h_{B}^{j}(I) \\
\label{reconnect4}
&&+  c_{A,j,1}^{\dagger}c_{A,j,1} h_{B}^{j}(N).
\end{eqnarray}
The terms (\ref{reconnect1}) and (\ref{reconnect2}) ensure that it is 
only when the target qubit $B$ enters the gate, that the control qubit 
$A$ can pass through with an identity operation.  If control qubit $A$ 
has value $0$ once it passes through to row $j$, then 
$c_{A,j,0}^{\dagger}c_{A,j,0}$ is non-zero and term 
(\ref{reconnect3}) subjects qubit $B$ to an identity operation.  
Term (\ref{reconnect4}) ensures that qubit $B$ is subjected to a NOT 
operation when qubit $A$ has value $1$ at row $j$.

%
%

\newcommand{\mycaption}[1]{\parbox{3in}{\small #1}}

\begin{figure}[h]
\centerline{\psfig{width=5.25cm,figure=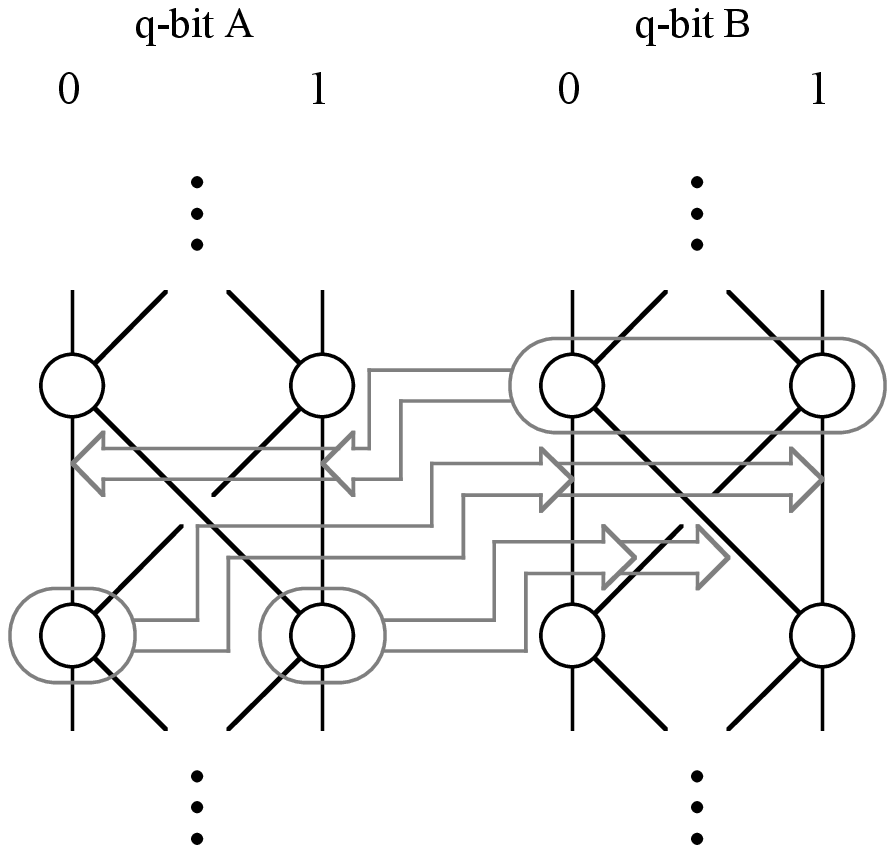}}
\vspace{0.125in}
\mycaption{FIG. 2.  Section of a two q-bit ground state 
quantum computer that implements a controlled NOT operation.  Arrows 
mark conditional tunneling paths.  }
\label{multi}
\end{figure}

In Fig.  2, we suggest how the controlled-NOT gate could be 
realized physically in a quantum-dot implementation.  The connection 
(\ref{reconnect1}-\ref{reconnect4}) proceeds through two-body 
interactions, in which the electric field of the electron in the 
control qubit influences the tunneling matrix elements and on-site 
energies in the target qubit and vice versa.

The generalization of the state development and construction of the 
Hamiltonian to operations involving multiple simultaneous CNOT 
operations is straightforward.  The incremental construction 
of the ground state and the projection property of equation 
(\ref{projectionprop}) guarantee that the state of the full quantum 
computer represents the full algorithm in that
\begin{equation}
\left| \Psi _{n} ^{N}\right> = 
{\cal U}_{N}{\cal U}_{N-1}\ldots{\cal U}_{1}
A_{N,0}P_{0} \left| \Psi_{n} ^{0}\right>.
\end{equation}
As in the single qubit computer, the initial state of the register 
$P_{0}\left| \Psi_{n}^{0}\right>$ is selected physically by biasing 
one of the initial stage quantum dots for each qubit.  This selects as 
input the binary representation of a number between $0$ and $2^{M}-1$.

At this point, we have outlined a functioning quantum computer with
both one and two qubit gates.  For purposes of illustration, we now
detail the specific implementation of a two-bit database search,
developed by Grover \cite{Grover1,Grover2} and described and
implemented in an NMR system by Chuang, Gershenfeld, and Kubinec
\cite{Chuang}.  Following the description of reference \cite{Chuang},
we note that the algorithm requires the operators $W$, $C$, and $P$,
where $W$ is a Walsh-Hadamard transform, $C$ is a conditional sign
flip, and $P$ is a different conditional sign flip.  These operators
can be implemented in the following manner.

The operation $W$ consists of a rotation by $\pi /4$ and a NOT for 
each qubit.  To achieve these single bit operations, we add two rows 
to each qubit, one for the rotation and one for the NOT.

The operation $C$ is a conditional sign flip which changes the sign of 
the state if and only if both qubits are in the 1 state.  This 
operation can be implemented by rotating the second bit by $\pi/4$, 
doing a NOT of the second bit if the first bit has a value of 1, and 
then rotating the second bit by $-\pi/4$.  Thus $C$ requires that we 
add three more rows to each qubit.

The other conditional sign flip, $P$, is achieved analogously.  It
requires an initial rotation of qubit $a=2$ by $-\pi/4$, a
controlled-NOT in which qubit $a=2$ suffers a NOT if qubit $a=1$ has a
0 value, and then a final rotation of qubit $a=2$ by $\pi/4$.
Thus, like $C$, $P$ requires three rows in each qubit.

From our analysis of the operations $W$, $C$, and $P$, it follows
that the most naive implementation of the full algorithm $WPWCW$ will
require 12 rows total, plus an initial input row.  This number can be
reduced by taking advantage of cancellations and other improvements.
 
In conclusion, we have presented a new ground state approach to 
quantum computation.  With this scheme, many of the technical problems 
associated with realizing a traditional quantum computer are 
circumvented, while different challenges arise.  Time-dependent or 
time-independent perturbations of the Hamiltonian could introduce 
errors into the calculation.  Static perturbations due to imperfect 
implementation of the requisite Hamiltonian will adversely influence 
the ground state.  Thus ground state quantum computation does not 
require time-dependent control of a system, but it does demand fine 
tunability of a static Hamiltonian.  Time-dependent perturbations such 
as thermal fluctuations are capable of exciting the system out of the 
ground state.  In a traditional quantum computation these fluctuations 
would lead to decoherence.  To the extent that such excitations can 
be quenched by large energy level spacings and low temperatures, they 
do not disturb ground state quantum computation.  

The implementation that we have proposed using quantum dot arrays may 
or may not be suitable for overcoming these sources of error.  While 
the quantum-dot implementation may seem technically challenging, it is 
encouraging to note that a classical computation scheme using coupled 
quantum dots has been implemented \cite{automata}.  Certainly, many 
other implementations could be envisioned; for example, the states 
$\left|0_i\right>$ and $\left|1_i\right>$ of a single qubit could take 
different locations in momentum space rather than different locations 
in real space.

We thank D. DiVincenzo for calling to our attention the ``cursor
Hamiltonian'' approach of references \cite{Feynman,Peres} and for
helpful comments on the manuscript.  We also thank A. E. Charman for
pointing out the research of reference \cite{automata}.

This work was supported by National Science Foundation grant
No. DMR-9520554; the Director, Office of Energy Research, Office of
Basic Energy Services, Materials Sciences Division of the
U.S. Department of Energy under Contract No. DE-AC03-76SF00098; and
the Office of Naval Research grant No. N000149610034.

\end{multicols} 

%

\end{document}